\documentclass[letter,traditabstract]{aa} 
\usepackage{hyperref}
\usepackage{epsf}
\usepackage{graphicx}
\usepackage{txfonts}
\usepackage{natbib}
\usepackage{longtable}

\def\ms{\hbox{\,m\,s$^{-1}$}}         
\def\m2s2{\hbox{\,m$^{2}$\,s$^{-2}$}} 
\def\kms{\hbox{\,km\,s$^{-1}$}}       
\def\vsini{\hbox{$v$\,sin\,$i$}}      
\def\Msun{\hbox{$\mathrm{M}_{\odot}$}}             
\def\Rsun{\hbox{$\mathrm{R}_{\odot}$}}
\def\Mjup{\hbox{$\mathrm{M}_{\rm Jup}$}}
\def\Rjup{\hbox{$\mathrm{R}_{\rm Jup}$}}

\def \sophie{SOPHIE}
\def \kepler{\emph{Kepler}}

\def\kepler{\emph{Kepler}}

\def \kms{km\,s$^{-1}$}
\def \1s{$1\,\sigma$}

\def \t0{T$_0$}

\def\Msun{\hbox{$\mathrm{M}_{\odot}$}}             
\def\Rsun{\hbox{$\mathrm{R}_{\odot}$}}
\def\Mjup{\hbox{$\mathrm{M}_{\rm Jup}$}}
\def\Rjup{\hbox{$\mathrm{R}_{\rm Jup}$}}
\def \415{KOI-415}

\begin{document}

\title{SOPHIE velocimetry of \kepler\ transit candidates \thanks{Based on observations collected  with the NASA \kepler\ satellite and with the {\it SOPHIE}  spectrograph on the 1.93-m telescope at Observatoire de Haute-Provence (CNRS), France.}}
            
\subtitle{IX. KOI-415\,b: a long-period, eccentric transiting brown dwarf to an evolved Sun}

\author{
C.~Moutou\inst{1}, A.~S.~Bonomo\inst{2},  G.~Bruno\inst{1},  G.~Montagnier\inst{3,4}, F.~Bouchy\inst{1}, J.~M.~Almenara\inst{1},  S.~C.~C.~Barros\inst{1},  M.~Deleuil\inst{1}, R.~F.~D\'iaz\inst{1},  G.~H\'ebrard\inst{3,4}, A.~Santerne\inst{5}
}

\institute{
Aix Marseille Universit\'e, CNRS, LAM (Laboratoire d'Astrophysique de Marseille) UMR 7326, 13388, Marseille, France
\email{Claire.Moutou@oamp.fr}
\and INAF - Osservatorio Astrofisico di Torino, via Osservatorio 20, 10025, Pino Torinese, Italy
\and Institut d'Astrophysique de Paris, UMR7095 CNRS, Universit\'e Pierre \& Marie Curie, 98bis boulevard Arago, 75014 Paris, France 
\and Observatoire de Haute-Provence, CNRS/OAMP, 04870 Saint-Michel-l'Observatoire, France 
\and Centro de Astrof\'isica, Universidade do Porto, Rua das Estrelas, 4150-762 Porto, Portugal}

\abstract{We report the discovery of a long-period brown-dwarf transiting companion of the solar-type star \415. The transits were detected by the \kepler\, space telescope. We conducted Doppler measurements using the SOPHIE spectrograph at the Observatoire de Haute-Provence. The photometric and spectroscopic signals allow us to characterize a 62.14 $\pm$ 2.69 \Mjup\, brown-dwarf companion of an evolved 0.94 $\pm$ 0.06 \Msun\, star in a highly eccentric orbit of $P$ = 166.78805 $\pm$ 0.00022 days and $e$ = 0.698 $\pm$ 0.002. The radius of KOI-415\,b is $0.79_{-0.07}^{+0.12}$ \Rjup, a value that is compatible with theoretical predictions for a 10 Gyr, low-metallicity and non-irradiated object. 
}


\date{Received TBC; accepted TBC}
      
\authorrunning{Moutou et al.}
\titlerunning{KOI-415\,b: a long-period, highly-eccentric transiting brown dwarf to an evolved Sun}

\offprints{\\
 \email{claire.moutou@oamp.fr}}

\maketitle

%

\section{Introduction}

Transiting brown dwarfs (BD) are much less common than transiting giant planets. Above 20 \Mjup, only eight such objects are known: CoRoT-3\,b \citep{deleuil}, KELT-1\,b \citep{siverd}, KOI-205\,b \citep{diaz}, the  young pair of eclipsing brown dwarfs  2M0535-05 \citep{stassun07}, WASP-30\,b \citep{triaud}, LHS 6343 C \citep{johnson}, and CoRoT-15\,b \citep{bouchy}. The precise measurement of their mass and radius offers important constraints for the transition region between giant planets and low-mass stars, to be compared with theoretical predictions as given by, for instance, \citet{burrows97}, \citet{chabrier00} and \citet{baraffe03}. If CoRoT-15\,b appears to be inflated, WASP-30\,b, LHS\,6343\,C and KOI-205\,b have a radius compatible with their age, as predicted by models. 2M0535-05\,A and B may have smaller radii than predicted, because of their activity \citep{stassun12}.

In this paper, we report the discovery and characterization of a new transiting brown dwarf in a long-period and highly-eccentric orbit: KOI-415\,b is one the $\sim$2300 transiting candidates identified by \kepler\, \citep{batalha}. The host star is a 14.11 magnitude star identified as KIC 6289650 or 2MASS-19331345+4136229. The transiting companion has been established as a brown dwarf by complementary observations with the SOPHIE radial-velocity instrument. KOI-415\,b is one of the few known \kepler\ Objects of Interest that lie in the brown dwarf domain and not in the planet domain (as KOI-205\,b), a type of false-positive scenario whose low-occurrence is favored by the "brown-dwarf desert", however. 

\section{Data}

\subsection{\emph{Kepler} photometry}
\label{kepler_photometry}
KOI-415 was observed by \emph{Kepler} 
with a temporal sampling of 29.4~min (long-cadence data) 
for 1239.8 days; short-cadence data (one point per minute) 
are not available for this target.
The $ \sim 59\,000$ raw simple-aperture-photometry measurements, which cover 
fourteen quarters, from Q1 to Q14, were downloaded from the
MAST archive\footnote{http://archive.stsci.edu/kepler/data\_search/search.php}.
The flux excess caused by background stars that contaminate
the photometric mask, as estimated by the \emph{Kepler} team 
for each quarter\footnote{http://archive.stsci.edu/kepler/kepler\_fov/search.php},
does not exceed $5\%$ of the total flux collected
by \emph{Kepler} and was subtracted from the raw data. 

\begin{figure}[h]
\centering
\includegraphics[width=5.5cm, angle=90]{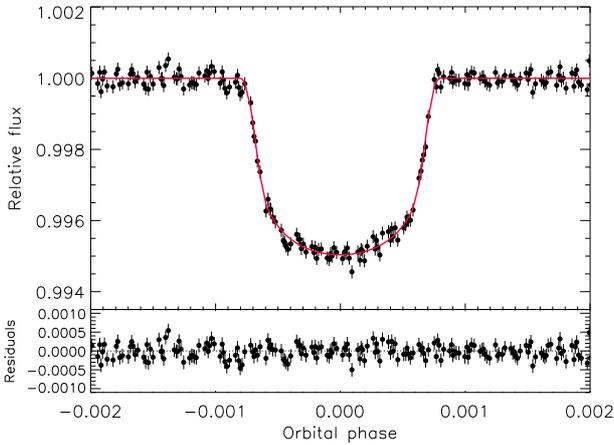}
\vspace{+0.5cm}
\caption{
\emph{Top panel}: phase-folded transit light curve of KOI-415. The red solid line
shows the transit model rebinned at the \emph{Kepler} sampling rate (see text for explanation). 
\emph{Bottom panel}: the residuals between the observations and the best-fit model.}
\label{tr_bestfit_fig}
\end{figure}

The light curve exhibits seven transits with a period of $166.8$~d, 
a depth of $\sim 0.5\%$, and a duration of $\sim 6$~h.
The median of the errors of the individual measurements 
is 185~ppm. Figure~\ref{tr_bestfit_fig} displays the phase-folded transit of KOI-415\,b.
 
No variations attributable to starspots
and faculae on the stellar photosphere are noticed, which 
indicates that the star is magnetically quiet.

\subsection{\sophie\ spectroscopy and velocimetry}

We have secured 14 measurements of \415\ from July 2012 to June 2013 
with the SOPHIE spectrograph at Observatoire de Haute-Provence. SOPHIE is an echelle optical spectrograph in a thermally controlled room, fed by a fiber link from the Coud\'e focus of the 1.93-m telescope. 
It covers the range 390-687 nm at a resolving power of 40,000 in high-efficiency (HE) mode; the acceptance on the sky is 3" diameter. The radial velocities (RV) were extracted with the SOPHIE pipeline 
by measuring the cross-correlation function (CCF) of the spectra with a numerical mask of a G2 star, following the method described in \citet{baranne}. The CCF was measured without the seven bluest orders which are dominated by noise. The G2 mask is preferred above the other F0 and K5 masks because it gives lower-noise measurements. The average RV uncertainty of the 14 measurements of \415\ is 31\ms. 

When the background was bright, the CCF in the neighboring sky fiber was subtracted from the stellar CCF before the stellar CCF was fitted by a Gaussian model. In four spectra out of the 14, the background was detected and the RV were corrected for the contamination, with offsets varying from 13\ms\ to 100\ms. The final measurements are listed in Table \ref{table:rv} together with the exposure time, signal-to-noise ratio per spectrum, and the bisector span of the CCF. 

The radial velocities exhibit a scatter of 2 \kms. Their variations are in phase with the \kepler\ photometric transit event and compatible with the highly eccentric orbit of a massive object, as plotted in Figure \ref{rv}. 

The bisector spans were used to check that the stellar CCF was not blended by an additional stellar spectrum, as is commonly done in transit-validation processes \citep{torres,bouchy09}. 
The bisector spans of \415\, do not vary significantly in our data set, and show a dispersion of 75$\pm$60\ms.



\section{Host star \label{sec.stellarparams}}

After RV corrections, all spectra that were only weakly affected by the moon reflected light were co-added. This resulted in a final spectrum with a signal-to-noise ratio of 120 per pixel at 5500\AA. The effective temperature $T_{\rm{eff}}$, surface gravity $\log g$, and metallicity $[\rm{Fe/H}]$ were estimated by using the versatile wavelength analysis package (VWA) (\citet{bruntt04} and \citet{bruntt10}). We selected 242 spectral lines of FeI and FeII and fitted them iteratively, until the derived abundances minimized the correlation with the respective equivalent width and excitation potential and the mean FeI and FeII abundances coincided.  $v_{macro}$ was fixed according to $T_{\rm{eff}}$ following \citet{bruntt10} to the value 2.4 \kms. $v_{micro}$ was fixed to 1.6\kms. \\
KOI-415 appears to be a slightly evolved solar-type star with  $T_{\rm{eff}}$ = 5810 $\pm$ 80 K, $\log g = 4.5 \pm 0.2$ and $[\rm{Fe/H}]$=-0.24 $\pm$ 0.11. We derived $\log g$ from the pressure-sensitive lines of Ca at 6122, 6162 and 6439 \AA~ and from the MgIb line at 5172 \AA, as well. The best fit was obtained from the Ca6122 and the Ca6162 lines ($4.34 \pm 0.26$ dex) and agrees with the result from the Fe lines. Abundances of elements presenting isolated lines were calculated. Using an estimated $B-V$ value of 0.58, the CCF parameters correspond to a projected rotational velocity of 2.8$\pm$1.5 \kms.  The spectroscopic $v\sin i$ was derived by fitting a set of isolated spectral lines to the corresponding synthetic profiles, which were convolved with different rotational profiles. Its low value ($1 \pm 1 \, \mathrm{km \, s}^{-1}$) is compatible within 1.2$\sigma$ with the CCF estimate. The low $v \sin i$ confirms the hypothesis that KOI-415 is an evolved star. Furthermore, no activity was observed through the CaII H and K lines. 
Using the spectroscopically determined \vsini\ and the stellar radius, the projected spin period of the star is 30 to 60 days.  The final parameters of the host star are derived after the transit modeling in the next section.

\section{Modeling the data and parameter estimation}
\label{modeling}

Before starting the combined analysis of \kepler\ and SOPHIE data,
a preliminary fit of the radial-velocity data
was performed with a Keplerian model 
by using the downhill simplex algorithm 
\citep{Pressetal92} and fixing the 
transit epoch and the orbital period to the values
determined from the photometry.
From the best solution, the uncertainties of SOPHIE
data were scaled to derive a reduced $\chi^2$ = 1. 
The scaling factor corresponds to an RV jitter of $39\ms$. 

The transits were normalized
by locally fitting a parabola to the 10~h intervals of the light curve before 
the ingress and after the egress of each transit.
The seventh transit, which occured at 
$\rm 2456078.870~BJD_{TBD}$, 
was not considered for our analysis
because a gap taking place before its ingress
prevented us from properly normalizing it.
Photometrically correlated noise was estimated 
following \citet{Pontetal2006} and \citet{Bonomoetal2012},
but its contribution was found to be negligible.

The eleven free parameters of our combined fit
are the transit epoch $T_{\rm 0}$, the orbital period $P$, 
the systemic RV $V_{\rm r}$, 
the RV semi-amplitude $K$, 
$\sqrt{e}~{\cos{\omega}}$ and  
$\sqrt{e}~{\sin{\omega}}$, where $e$  is the eccentricity and
$\omega$ the argument of periastron, 
the transit duration $T_{\rm 14}$,
the ratio of the brown dwarf to stellar radii $R_{\rm b}/R_{*}$, 
the inclination $i$ between the orbital plane and the plane of the sky,
and the two limb-darkening coefficients $u_{+}=u_{a}+u_{b}$
and $u_{-}=u_{\rm a}-u_{\rm b}$. 

The transit fitting was carried out using the model 
of \citet{Gimenez06, Gimenez09} and a denser temporal 
sampling $\delta t_{\rm model}=\delta t/5$, 
to overcome the problem of the coarse long-cadence sampling 
that distorts the transit shape \citep{Kipping10}. 
The $\chi^2$ of each trial model was then computed by binning the model 
samples at the \emph{Kepler} sampling rate $\delta t$. 
An initial combined fit was performed by using the 
algorithm AMOEBA \citep{Pressetal92} and changing the initial
values of the parameters with a Monte Carlo method to 
properly explore the parameter space. 

The posterior distributions of our free parameters 
were determined by means of a 
differential evolution Markov chain Monte Carlo (DE-MCMC) method 
which is an MCMC version of the genetic algorithm 
\citep{TerBraak2006, Eastmanetal2013}. 
Twenty-two chains, which is twice the number of the
fitted parameters, were simultaneously run after starting 
at different positions in the parameter space 
but reasonably close to the best solution 
previously found with AMOEBA. 
The jumps for a current chain in the parameter space 
were determined from the other chains according
to the prescriptions given by \citet{TerBraak2006}.
The Metropolis-Hastings algorithm was used 
to accept or reject a proposal step for each chain.
Uniform priors were implicitly imposed for all parameters 
except for the limb-darkening coefficients,
for which Gaussian priors were considered, 
that is, $N_{u_{+}}(0.7, 0.07)$ and $N_{u_{-}}(0.15, 0.14)$, 
where the most probable values were determined
by the exploratory fit with AMOEBA.

Our DE-MCMC stopped when the convergence of the
chains was achieved according to \citet{Ford2006}, that is,
when the Gelman-Rubin statistics was lower than 1.01 
\citep{Gelmanetal2004} and the number of 
independent draws was greater than 1000.
Steps belonging to the ''burn-in'' phase, 
which were identified following \citet{Knutsonetal2009}
and \citet{Eastmanetal2013},
were discarded.

The medians of the posterior distributions of the 
fitted and derived parameters 
and their $34\%$ intervals are quoted in 
Table \ref{table:starplanet_param_table} as the final values 
and their $1\sigma$ uncertainties, respectively.

\begin{figure}[t]
\begin{center}
\includegraphics[width=0.9\columnwidth]{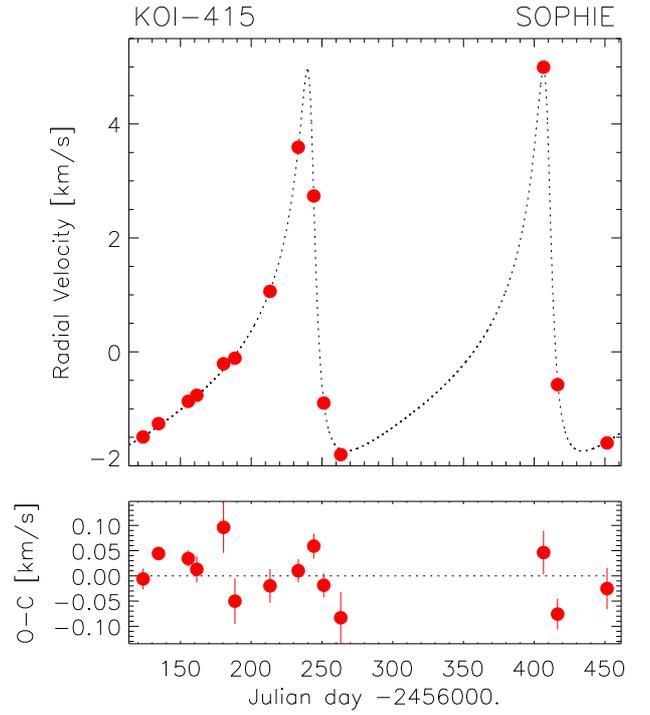}\hfill
\caption{\label{rv} 
Radial velocities of KOI-415 obtained by SOPHIE as a function of time. The bottom plot shows the residuals after subtracting the model. }
\end{center}
\end{figure}

The RV observations carried out with SOPHIE and the solution of the Keplerian fit derived from our combined fit are shown in Fig.~\ref{rv}. The scatter of the residuals is 71 m/s. Figure \ref{tr_bestfit_fig} displays the phase-folded transit of KOI-415\,b and the transit model. The photometric residuals show an rms of 177~ppm, consistent with the median error of the data points (185~ppm). The rms over 3 hours is 72~ppm, in agreement with the combined differential photometric precision of   $ \sim 85 \rm ppm$ \citep{Gillilandetal2011} provided by the \emph{Kepler} team. 

\onltab{1}{
\begin{table}[t]
\caption{Radial velocity measurements for \415 \label{table:rv}. Asterisks show the measurements corrected for the background light.}
\begin{tabular}{l l l c c c}
\hline
\hline
\noalign{\smallskip}
$\rm BJD_{UTC}$     & RV          & $\sigma_{RV}$ & Bis  & T$_{exp}$ & S/N/pix    \\
-2456000  &(\kms) & (\kms) & (\kms) & (s)       & (550 nm)\\
\hline
\noalign{\smallskip}
123.5860& -2.903& 0.021& -0.052& 2700&14\\
134.4912& -2.669& 0.012& -0.006& 3600& 22\\
155.4896& -2.279& 0.016& 0.021& 3600& 19\\
161.5256& -2.169& 0.027& 0.011& 1521& 13\\
180.4504& -1.594& 0.052& -0.116& 900& 10\\
188.4825& -1.533& 0.047& 0.141& 900& 9\\
213.3051& -0.349& 0.034& -0.103& 1940& 13\\
233.2643& 2.172*& 0.024& 0.038 &1800& 16\\
244.2319& 1.371*& 0.032& -0.097& 1210& 12\\
251.2602& -2.313*& 0.024& 0.027& 1800& 16\\
263.2605& -3.181& 0.054& 0.074& 1800& 10\\
406.5797& 3.569*& 0.046& 0.0013 &2700& 10\\
416.5095& -1.985& 0.031& 0.052& 1503& 14\\
451.4736& -2.987& 0.041& 0.046& 2202& 11\\
\noalign{\smallskip}
\hline 
\end{tabular}
\end{table}
}

The stellar density derived from the transit fitting and the Yonsey-Yale evolutionary
tracks \citep{Demarqueetal2004} for the effective temperature and metallicity of KOI-415
indicate that with a mass $M_\star =0.94 \pm 0.06~\Msun$, a 
radius $R_\star =1.25_{-0.10}^{+0.15}~\Rsun$, and age of $10.5 \pm 2.2$ Gyr at $1\sigma$,
this star is slightly evolved. This is compatible with the absence of activity-related
features in the \emph{Kepler} light curve.
The photometric log\,$g=4.22_{-0.09}^{+0.06}$~dex agrees with 
the spectroscopic value, which is 
more uncertain.

The derived coefficients of the quadratic limb-darkening
law $u_{\rm a}=0.42 \pm 0.05$ and $u_{\rm b}=0.28 \pm 0.07$ 
agree very well with the values determined by \citet{Sing10}
using Kurucz stellar models \citep{Kurucz93} and the \emph{Kepler} 
bandpass\footnote{$\rm  http://vega.lpl.arizona.edu/singd/David\_Sing/Limb\_Darkening.html$}, 
after linearly interpolating at the $T_{\rm eff}$, log\,$g$ and metallicity of the star,
that is, $u_{\rm a}=0.36 \pm 0.02$ and $u_{\rm b}=0.28 \pm 0.01$.

According to the determined stellar parameters, the mass and radius of the 
brown dwarf KOI-415\,b are $M_{\rm b}= 62.14  \pm 2.69~\Mjup$ and 
$R_{\rm b}=0.79_{-0.07}^{+0.12}~\Rjup$. The 
corresponding density is $157.4_{-52.3}^{+51.4}\, \rm g\;cm^{-3}$. 
Its very large error is dominated by the uncertainty on the brown dwarf radius.

\begin{table}
\centering
\caption{KOI-415 system parameters.}            
\begin{minipage}[t]{13.0cm} 
\setlength{\tabcolsep}{1.0mm}
\renewcommand{\footnoterule}{}                          
\begin{tabular}{l l}        
\hline\hline                 
\multicolumn{2}{l}{\emph{Fitted system parameters}} \\
\hline
BD orbital period $P$ [days] & 166.78805 $\pm$ 0.00022 \\
BD transit epoch $T_{ \rm 0} [\rm BJD_{TDB}-2454900$] & 178.14187 $\pm$ 0.00063 \\
BD transit duration $T_{\rm 14}$ [h] & $6.01_{-0.07}^{+0.11}$ \\
Radius ratio $R_{\rm p}/R_{*}$ & $0.0649_{-0.0013}^{+0.0017}$  \\
Inclination $i$ [deg] & $89.31_{-0.38}^{+0.40}$ \\
Limb-darkening coefficient $u_{+}$  & 0.70 $\pm$ 0.05 \\
Limb-darkening coefficient $u_{-}$  &  0.13 $\pm$ 0.12  \\
$\sqrt{e}~\cos{\omega}$ & 0.592 $\pm$ 0.005 \\
$\sqrt{e}~\sin{\omega}$ & 0.590 $\pm$ 0.005 \\
Radial velocity semi-amplitude $K$ [\kms] & 3.346 $\pm$ 0.021 \\
Systemic velocity  $V_{\rm r}$ [\kms] & -1.492 $\pm$ 0.011 \\
& \\
\multicolumn{2}{l}{\emph{Derived orbital parameters}} \\
\hline
Orbital eccentricity $e$  &  0.698 $\pm$ 0.002 \\
Argument of periastron $\omega$ [deg] & 44.9 $\pm$ 0.6 \\ 
& \\
\multicolumn{2}{l}{\emph{Derived transit parameters}} \\
\hline
$a/R_{*}$ & $100.0_{-10.1}^{+7.5}$ \\
Stellar density $\rho_{*}$ [$ \rm g\;cm^{-3}$] & $0.68_{-0.18}^{+0.16}$\\
Impact parameter $b$& $0.41_{-0.23}^{+0.16}$ \\ 
Limb-darkening coefficient $u_{a}$  & 0.42 $\pm$ 0.05 \\
Limb-darkening coefficient $u_{b}$  &  0.28 $\pm$ 0.07  \\
& \\
\multicolumn{2}{l}{\emph{Atmospheric parameters of the star}} \\
\hline
Effective temperature $T_{\rm{eff}}$[K] & 5810 $\pm$ 80 \\
Spectroscopic surface gravity log\,$g$ [cgs]&  4.5 $\pm$  0.2 \\
Photometric surface gravity log\,$g$ [cgs] ~$^b$ &  $4.22_{-0.09}^{+0.06}$ \\
Metallicity $[\rm{Fe/H}]$ [dex]& -0.24 $\pm$ 0.11 \\
Stellar rotational velocity $V \sin{i_{*}}$ [\kms] & 1.0 $\pm$ 1 \\
Spectral type & G0IV \\
& \\
\multicolumn{2}{l}{\emph{Star and companion physical parameters}} \\
\hline
Stellar mass [\Msun]~$^a$ &  0.94 $\pm$ 0.06 \\
Stellar radius [\Rsun] ~$^a$ &  $1.25_{-0.10}^{+0.15}$  \\
BD mass $M_{\rm b}$ [\Mjup ]  &  62.14  $\pm$ 2.69 \\
BD radius $R_{\rm b}$ [\Rjup]  &  $0.79_{-0.07}^{+0.12}$ \\
BD density $\rho_{\rm b}$ [$\rm g\;cm^{-3}$] &  $157.4_{-52.3}^{+51.4}$ \\
BD surface gravity log\,$g_{\rm b }$ [cgs] &  $5.39_{-0.11}^{+0.08} $ \\
Age $t$ [Gyr]~$^a$  & 10.5 $\pm$ 2.2 \\
Orbital semi-major axis $a$ [AU] & 0.593 $\pm$ 0.013 \\
Orbital distance at periastron $a_{\rm per}$ [AU] & 0.179 $\pm$ 0.004 \\
Orbital distance at apastron $a_{\rm apo}$ [AU] &  1.006 $\pm$ 0.021 \\
\hline       
\hline       
\vspace{-0.5cm}
\label{table:starplanet_param_table}  
\end{tabular}
\end{minipage}
\end{table}

\section{Discussion}

\begin{figure}[t!]
\begin{center}
\includegraphics[width=1.05\columnwidth]{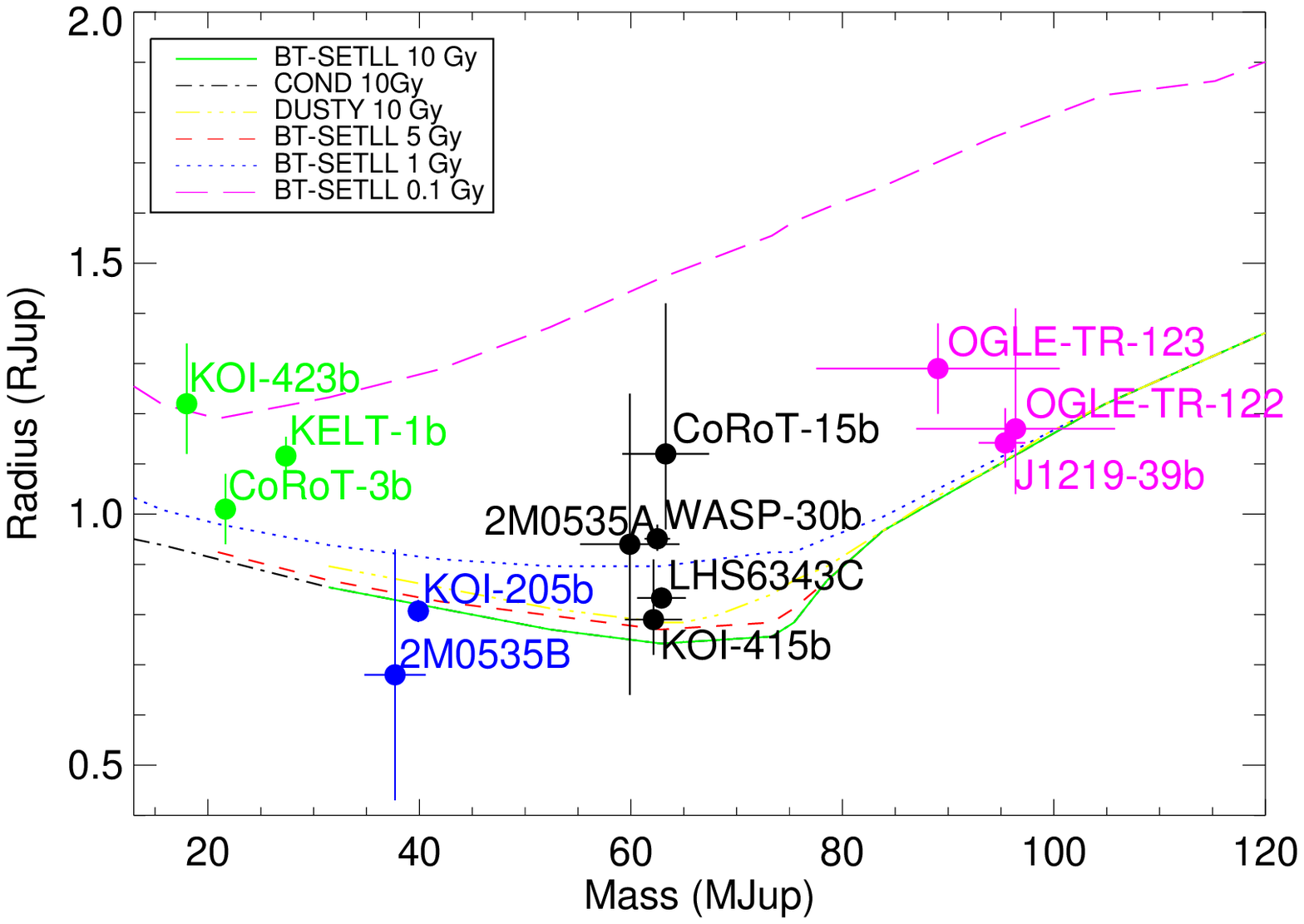}
\includegraphics[width=1.05\columnwidth]{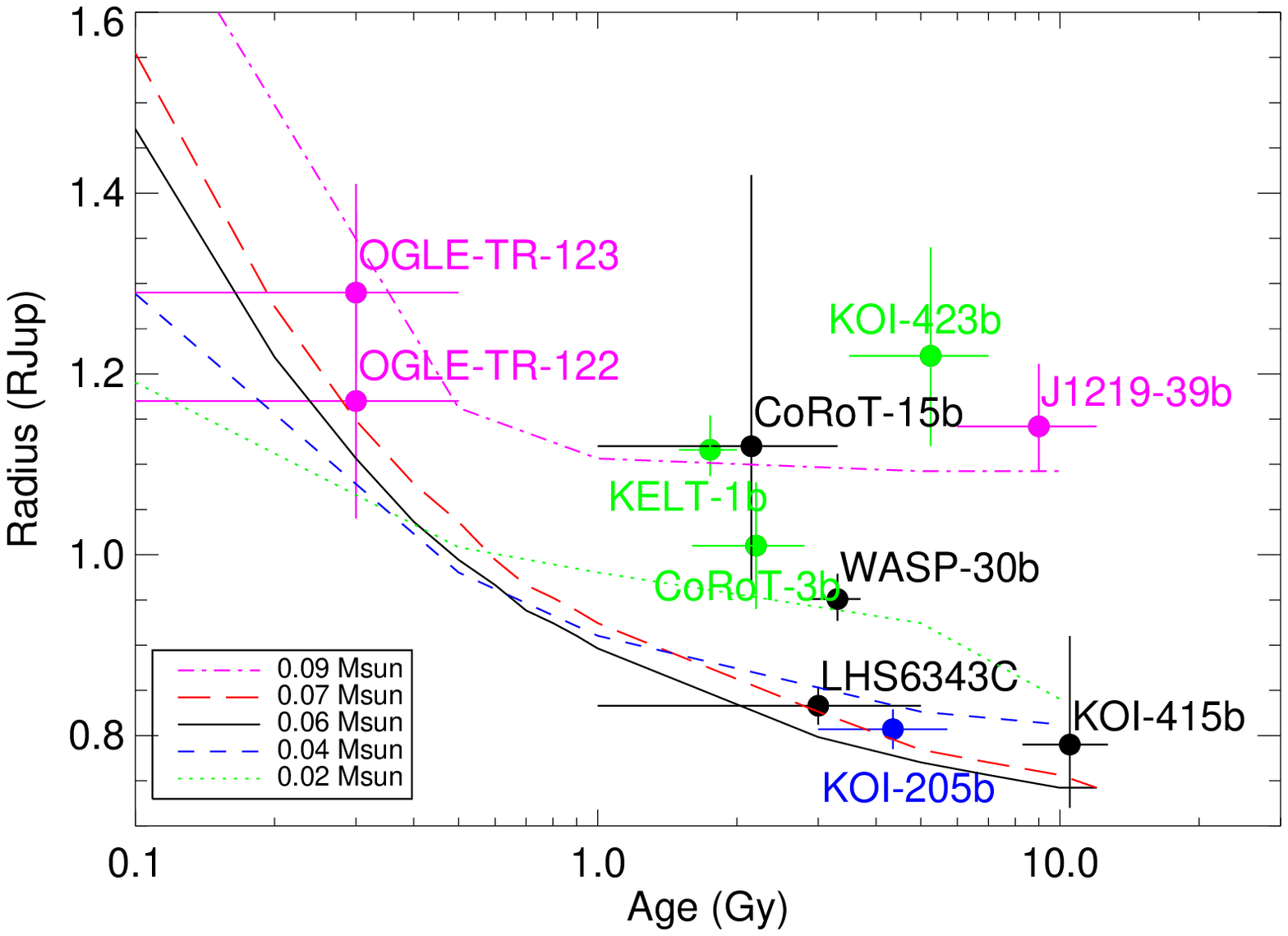}
\caption{\label{fig.mass_radius} \emph{Top panel}: The mass-radius diagram in the transition domain between brown dwarfs and stars. Isochrones for 10, 5, 1, and 0.1 Gyr are shown for comparison. \emph{Bottom panel}: the radius as a function of system's age for objects from 15 to 100 Jupiter masses. BT-SETTL isochrones are shown for 0.02 to 0.09 \Msun. Color symbols indicate the mass range: 15-25 \Mjup\ (green), 37-40 \Mjup\ (blue), 59-65  \Mjup\ (black), and 89-97 \Mjup\ (pink).}
\end{center}
\end{figure}

Figure \ref{fig.mass_radius} (top) shows the mass and radius of KOI-415\,b and other similar objects in the transition domain between giant planets and low-mass stars. They are compared with values predicted by isochrones\footnote{http://phoenix.ens-lyon.fr/Grids} for models of isolated BDs developed by \citet{chabrier00}, \citet{allard01}, and \citet{baraffe03}. With its long period (166.8 days), KOI-415\,b more closely resembles  the isolated BD as modeled in the tracks than the short-period companions more easily found by transiting surveys.
The radius of KOI-415\,b perfectly fits the predicted radius for a system age of 10 Gy and a non-metallic system with a cloud-free atmosphere \citep{burrows97,baraffe03,burrows11}. It shows that long-period BDs are adequately supported by partial electron degeneracy physics with a quantitatively correct equation of state. Brown dwarfs such as KOI-415\,b are among the smallest ones in size, corresponding to the evolved state of its system, its low metallicity and low irradiation. It is also the oldest characterized BD (Figure \ref{fig.mass_radius} (bottom)).

The models predict an effective temperature of 950K for KOI-415\,b. With an eccentricity of about 0.7 and period of 166.8 days, the temperature of the dayside of the brown dwarf likely increases by as much as 400K along its orbit, from periastron to apastron due to changing stellar irradiation. It could be interesting to investigate, with dedicated modeling, the impact on the internal structure this seasonal effect may have. 

Finally, the speckle imaging of the field around KOI-415 shows the existence of a nearby companion with a magnitude difference 2.9 in the optical, at 1.93" from the main target\footnote{https://cfop.ipac.caltech.edu/}. If it were bound to the main star, this companion would be an M1 dwarf at about 1600 AU orbital distance. The existence of such a stellar companion could then explain the high eccentricity of the brown dwarf KOI-415\,b. Doppler measurements collected over a few years timescale could be sufficient to unveil their relationship; a background star is also a possibility.

\begin{acknowledgements}
ASB gratefully acknowledges support through INAF/HARPS-N fellowship. We thank the staff at Haute-Provence Observatory. We acknowledge the PNP of CNRS/INSU, and the French ANR  for their support. RFD is supported by CNES. AS acknowledges the support by the European Research Council/European Community under the FP7 through Starting Grant agreement number 239953.

\end{acknowledgements}

\end{document}